\documentclass[twocolumn,amssymb, nobibnotes, floatfix, aps, prb]{revtex4-1}
\usepackage{hyperref}
\usepackage[utf8]{inputenc}
\usepackage[english]{babel}
\usepackage{amssymb}
\usepackage{amsmath}
\usepackage{graphicx}
\usepackage{color}
\usepackage{physics}
\usepackage{float}

\renewcommand{\theHequation}{M.\arabic{equation}}
\renewcommand{\theHfigure}{M\arabic{figure}}

\renewcommand{\vec}[1]{\boldsymbol{\mathbf{#1}}}
\begin{document}
\title{Electrostatic effects,  band distortions and superconductivity in twisted graphene bilayers}
\author{Francisco Guinea$^{1,2}$}\email{Francisco.Guinea@manchester.ac.uk}
\author{Niels R. Walet$^2$} \email{Niels.Walet@manchester.ac.uk}

\affiliation{$^1$Imdea Nanoscience, Faraday 9, 28015 Madrid, Spain}
\affiliation{$^2$School of Physics and Astronomy, University of Manchester, Manchester, M13 9PY, UK}


\begin{abstract}
Bilayer graphene twisted by a small angle shows a significant charge modulation away from neutrality, as the charge in the narrow bands near the Dirac point is mostly localized in a fraction of the Moir\'e unit cell. The resulting electrostatic potential leads to a filling dependent change in the low energy bands, of a magnitude comparable, or larger than the bandwidth. These modifications can be expressed in terms of new electron-electron interactions, which, when expressed in a local basis, describe electron assisted hopping terms. These interactions favor superconductivity at certain fillings.
\end{abstract}

\maketitle
The discovery of strong repulsive interactions and superconductivity in twisted graphene bilayers \cite{Ketal17,Cetal18a,Cetal18b} (see also previous experimental results~\cite{Letal11b,Letal11} and recent work~\cite{Hetal18d,Yetal18}) has lead to an increased interest in the role of the electron-electron interaction in these systems. Previously, extensive experimental studies had been carried out in graphene superlattices on a BN substrate, see for example Refs.~\cite{Yetal11,Petal13,Hetal13,Detal13}. The mismatch in lattice spacing between graphene and BN limits the maximum wavelength of the Moir\'e lattice. As the lattice constant in the two graphene layers in a twisted bilayer is the same, the periodicity of the Moir\'e structure diverges at small twist angles \cite{LPN07,M10,BM11,M11,LPN12}. For sufficiently small angles almost flat bands arise near the charge neutrality point \cite{TMM10,BM11}. The effects of the intrinsically small interaction effects in graphene are expected to be enhanced for those `magic' angles where the width of the low energy bands is smallest. Novel magnetic phases become possible when the lowest band is half filled \cite{GLGS17}. Layer dependent strains can also lead to Moiré structures and narrow bands \cite{SGG12,Hetal18b}.

In the following we consider the effect of the long-range Coulomb interaction on the bands nearest to the neutrality point, as function of their filling. 
The occupation of the low energy bands in a twisted bilayer graphene leads to inhomogeneous electrostatic potentials of order of magnitude $e^2 / ( \epsilon L_M )$, where $L_M$ is the length of the Moir\'e lattice unit, and $\epsilon$ is the dielectric constant of the environment.  We expect this interaction to be comparable, or larger, than the band splitting, and thus it will give rise to a significant distortion of the bands. We analyze the screening of the Coulomb interaction within the Hartree-Fock approximation. As this is a variational method, it can be reliably used to treat screening effects, even when the interaction energy is comparable to the kinetic and crystal field contributions. 


We find that the electrostatic potential due to the inhomogeneous charge distribution in the Moir\'e unit cell gives rise to significant distortions in the band. Using general properties  of the Wannier functions which describe this band \cite{PZVS18b,Ketal18b,KV18b}, we identify new local, filling dependent, couplings which change the band dispersion away from the neutrality point. Finally, we discuss the relation between these interactions, which can be defined as electron assisted hopping terms, and superconductivity. Our work presents a rather different angle, and thus complements the  extensive corpus of work on the role of local interactions \cite{XB18b,GZFS18b,BV18,Detal18,YF18b,PCB18,YV18,HYL18,XLL18b,TCSS18}. We will not consider the role of lattice relaxation\cite{NK17}, since this  is a complex problem and will be discussed separately. We also do not address the role of the electron-phonon interaction on superconductivity, see Ref.~\cite{WMM18,BWB18}.

{\it Estimates of the electron-electron interactions.}
We evaluate electron-electron interactions in the low energy bands of a twisted Moir\'e bilayer with twist angle $\theta$ to leading powers in $d / L_M$, where $d \approx 2.4\,\text{\AA }$ is the lattice unit of graphene, and $L_M = d / ( 2 \sin ( \theta / 2 ) )\approx d / \theta$ is the Moir\'e lattice unit. The charge distribution of the states in the bands near the neutrality point  is mostly concentrated in regions with $AA$ stacking. This inhomogeneous charge leads to electrostatic potentials of  strength $V_{\text{Coulomb}} \sim e^2 / ( \epsilon L_M )$. For an angle of the order $\theta \sim 1^\circ$, $\epsilon \approx 4$ and a Moir\'e pattern of wavelength $L_M \sim 10 - 20$ nm, we obtain a Coulomb potential in the order of $50\, \text{meV}$. 

The overlapping $\pi$ orbitals of the carbon atoms also generate an intra-atomic Hubbard interaction, $U \sim e^2 / d$, which prevents double occupancy. This potential, when projected onto wavefunctions of extensions similar to $L_M$, generates an effective interaction $U_{\text{eff}} \sim ( U N_M^{-2} )\,N_M$, where $N_M \approx ( L_M / d )^2$ is the number of atoms within the Moir\'e lattice unit, and the factor $N_M^{-2}$ originates from the normalization of the wavefunctions. As a result, $U_{\text{eff}} \sim ( e^2 d ) / L_M^2\sim 10^{-2} \times V_{\text{Coulomb}} \sim 1\, \text{meV}$. Thus, the intra-atomic Hubbard term is comparable to the electron bandwidth $W$ at the magic angles where $W \sim 2 - 10$ meV, and the long-range Coulomb interaction can be  larger. In the following, we only consider the effect of the long range part of the Coulomb interaction. A detailed analysis of the short range Hubbard repulsion can be found in Ref.~\cite{GLGS17}.

\begin{figure}
\begin{center}
\includegraphics[width=3in]{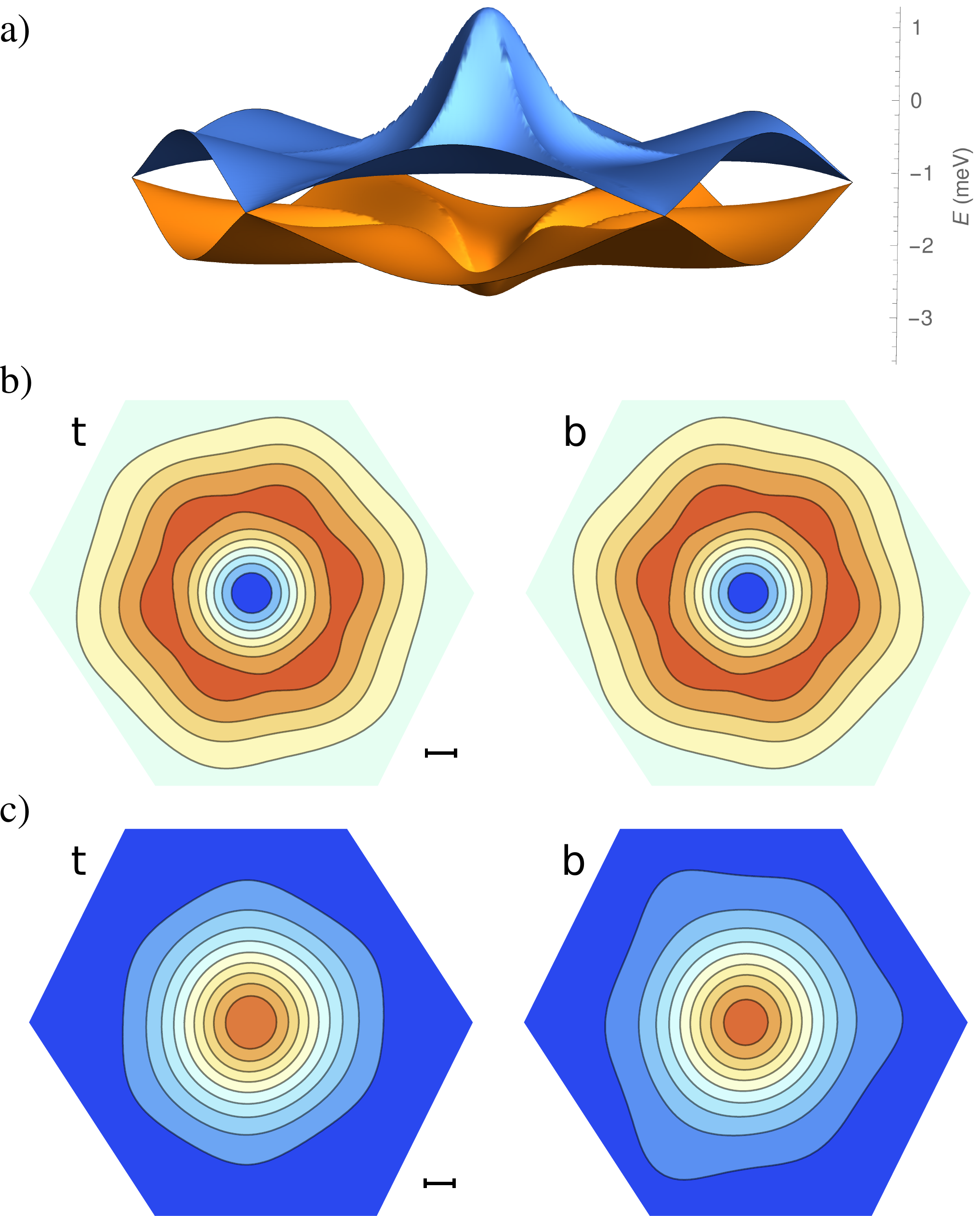}
\end{center}
\caption{a) Low energy bands of a twisted bilayer with twist angle $\theta = 1.05^\circ$ plotted over the full Brillouin zone (see also Fig.~\ref{fig:continuum}). b) Charge densities of states at the $\Gamma$ point in the Brillouin Zone (t for top layer; b for bottom). c) The charge density at the $K$ point summed over the two degenerate states. The black bars show a length of $1\,\mathrm{nm}$.}
\label{fig:dens}
\end{figure}

{\it Band structure and charge distribution.}
For small twist angles  the wavelength of the Moir\'e superlattice is much larger than  the graphene unit cell. It therefore makes sense to analyze  the band structure using the continuum approximation, see Refs.~\cite{LPN07,M10,BM11,LPN12}. We specifically use the parametrization of Ref.\ \cite{Ketal18b} for a Moir\'e pattern with a twist angle $\theta \approx 1.05^\circ$. Numerical values are given  in the Supplementary Information, Section 2. The bands, and the charge density at high symmetry points in the Brillouin zone are shown in Fig.~\ref{fig:dens}. As extensively discussed in the literature, the charge density associated to states at the edges of the Brillouin Zone, the $K$ and $M$ points, is concentrated in regions with $AA$ stacking. On the other hand, at the center of the Brillouin Zone, the $\Gamma$ point, the charge density vanishes in these $AA$ regions, and is much more homogeneously distributed throughout the rest of the Moir\'e cell. This asymmetry between charge densities in different regions of the Brillouin Zone has been pointed out previously in Ref.\ \cite{RM18}. As the bands in the two valleys are related by time reversal symmetry, we will consider in most of the following analysis a single layer, taking into account the fourfold degeneracy when necessary.

{\it Hartree potential.}
The Hamiltonian of  twisted bilayer graphene breaks electron-hole symmetry, and charge inhomogeneities can be expected, even at half filling. These fluctuations are expected to be small as, for low twist angles, the Moir\'e pattern can be decomposed into regions with $AA$, $AB$ and $BA$ stacking, which are locally neutral. In the following we neglect these charge inhomogeneities. The charge distribution which arises away from the neutrality point in inhomogeneous, and it is peaked in the $AA$ regions, so that a change in the occupancy of these bands will lead to charge fluctuations with the periodicity of the Moir\'e unit cell, $L_M$, as confirmed by numerical calculations described below. Fluctuations of the charge density and electrostatic potential are dominated by the first star of reciprocal lattice vectors, the six points with $| \vec{\bf G} | = ( 4 \pi ) / ( \sqrt{3} L_M )$. 

If we consider partially filled conduction bands, their contribution  to the charge density can be parametrized  by a dimensionless number,  $\rho_G$, a complex quantity which encodes the symmetric (real) and antisymmetric (imaginary) parts of the charge density. From the value of $\rho_G$ we can easily find the electrostatic potential in Fourier space, $v_{G} = V_0 =  v_{\text{Coulomb}} ( \vec{ G} ) \rho_G$, and $v_{\text{Coulomb}} ( \vec{ G} ) =(  2 \pi e^2 ) / ( \epsilon | \vec{ G} | )$. The amplitude of the fluctuations of the potential in real space are given by $V = ( 2 \pi e^2 \rho_G ) / ( \epsilon  \Omega | \vec{ G} | )$, where $\Omega = L_M^2 \sqrt{3} / 2$ is the area of the Moir\'e unit cell. We finally obtain $V = ( e^2 \rho_G ) / ( \epsilon L_M )$. In the following, we neglect the dependence of $\rho_G$ on sublattice and layer, as both vary over length scales of order $L_M$. The value of $\rho_G$ depends on the extent of the wavefunctions in momentum space, and it is bounded by $| \rho_G | < 4$, as it is a cross product of amplitudes times a spin-layer degeneracy factor $4$,
\begin{align}
\rho_G &= \bar{\rho}_G + \delta \rho_G = \bar{\rho}_G +4 \sum_{\vec{ k}}  \sum_{\vec{ G}'}  a^*_{\vec{ k} + \vec{ G} + \vec{ G'}} a_{\vec{ k}  + \vec{ G'}}.
\end{align}
Here $\bar{\rho}_G$ is a a constant which takes into account the contribution to the total density from all the bands not included in the calculation. 
We fix its value by imposing an homogeneous state at charge neutrality,  $\rho_G ( n_0 ) = 0$, where $n_0$ is the density at half filling. The coefficient $a_{\vec{ k} + \vec{ G}}$ is the amplitude that the wavefunction of band state $\vec{ k}$ is represented by a Bloch state with momentum $\vec{ k} + \vec{ G}$, and a sum over sublattice and layer indices is omitted. The wavefunctions of the lowest bands in twisted bilayer graphene are delocalized in momentum space, and we find $0.4 \lesssim | \delta \rho_G | \lesssim 4$. In all cases considered, ${\rm Im} ( \rho_G ) \lesssim 0.1 {\rm Re} ( \rho_G )$. This imaginary part arises from the small breaking of the symmetry between the $AA$ and the $AB$ and $BA$ regions in the Hamiltonian \cite{Ketal18b}. We therefore describe the Hartree potential in terms of its real part only.

We approximate the fully self consistent Hartree potential by describing it in terms of the six shortest reciprocal lattice vectors. Our calculations indicate that the next set of higher Fourier components are an order of magnitude smaller. The amplitude of the Hartree potential, $V_H$, determines the parameters of the Hamiltonian. As function of these parameters and the filling, $n$, we determine the charge density,  $\rho_G$.  The self consistency equation for the Hartree potential  can be written as
\begin{align}
V_H ( n ) &=  \frac{2 \pi e^2}{\epsilon | \vec{ G}  |} \rho_G ( V_H , n )\equiv  V_0 \times \rho_G ( V_H , n ).
\label{vh_selfc}
\end{align}
where $n$ is the band filling.

\begin{figure}
\begin{center}
\includegraphics[width=3in]{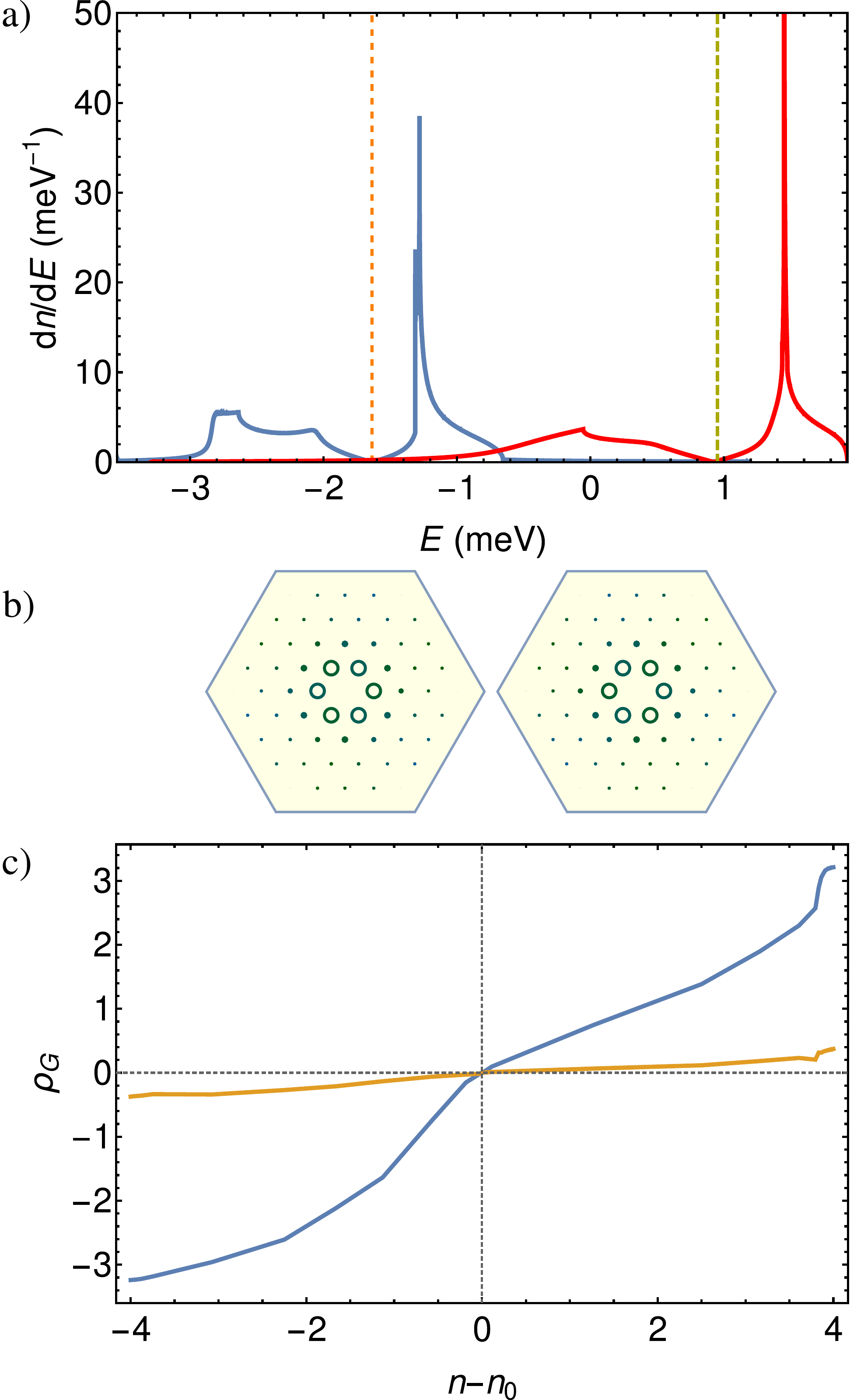}  \
\end{center}
\caption{a): Dependence of the density of states on the value of the Hartree potential for a twisted bilayer graphene with a twist angle $\theta = 1.05^\circ$. The blue line is for $V_H=0$, the red one for $V_H=1\,\text{meV}$. The vertical scale is truncated: the red peak extends more than twice as high as shown. The dotted lines are the Fermi energies at neutrality. b) Fourier transform of the charge density in the lowest state above the Fermi energy, with the dominant peak at zero momentum suppressed. The area of the circles is equal to the magnitude of the Fourier components, for each of the two layers. The momentum distributiion is concentrated at the six lowest wavevectors, $\vec{G}_i$, c) The value of the charge density at wavevector $\vec{G}_1$ as a function of the filling relative to charge neutrality. The blue line gives the real part, and the yellow line gives the imaginary part.}
\label{fig:continuum}
\end{figure}

In Fig.~\ref{fig:continuum}a we show the DOS for the original problem and also for a Hartree potential of amplitude $V_H = 1$ meV. The Hartree potential distorts the bands  considerably. The Fourier transform of the charge as function of band filling, shown in Fig.~\ref{fig:continuum}b, looks very similar for both cases--the difference is just in a normalization factor, and we only give the example for $V_H=0$. 

\begin{figure}[htb]
\begin{center}
\includegraphics[width=2.8in]{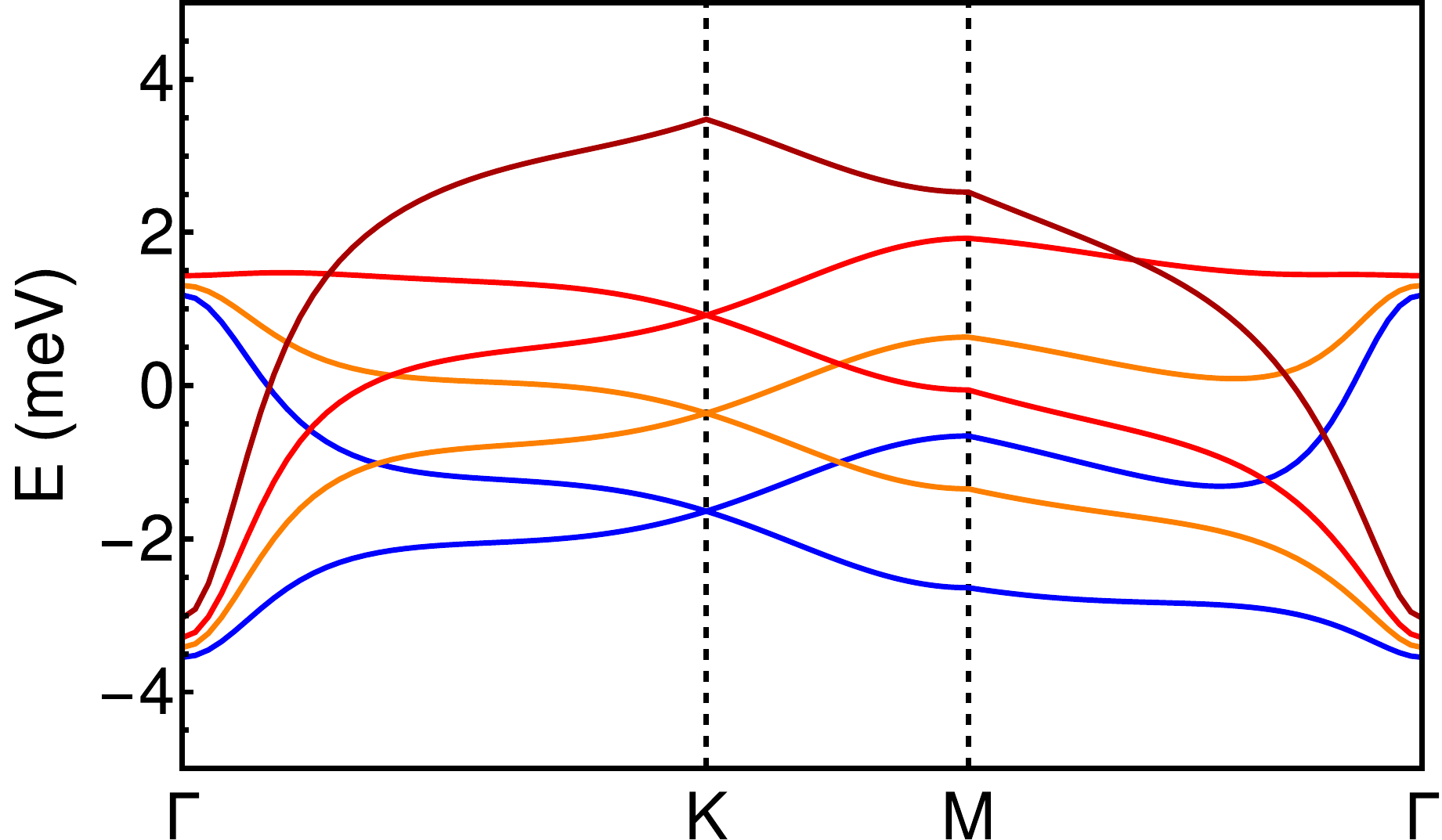}
\end{center}
\caption{Bands calculated for different electrostatic potentials. The blue lines are the bands at charge neutrality, the green lines are at $V_H=-0.5\,\mathrm{meV}$; the orange, the red and the dark-red lines are at  $V_H = 0.5 , 1$ and $2$ meV, respectively.}
\label{fig:comp_hartree}
\end{figure}

\begin{figure}
\begin{center}
\includegraphics[width=2.5in]{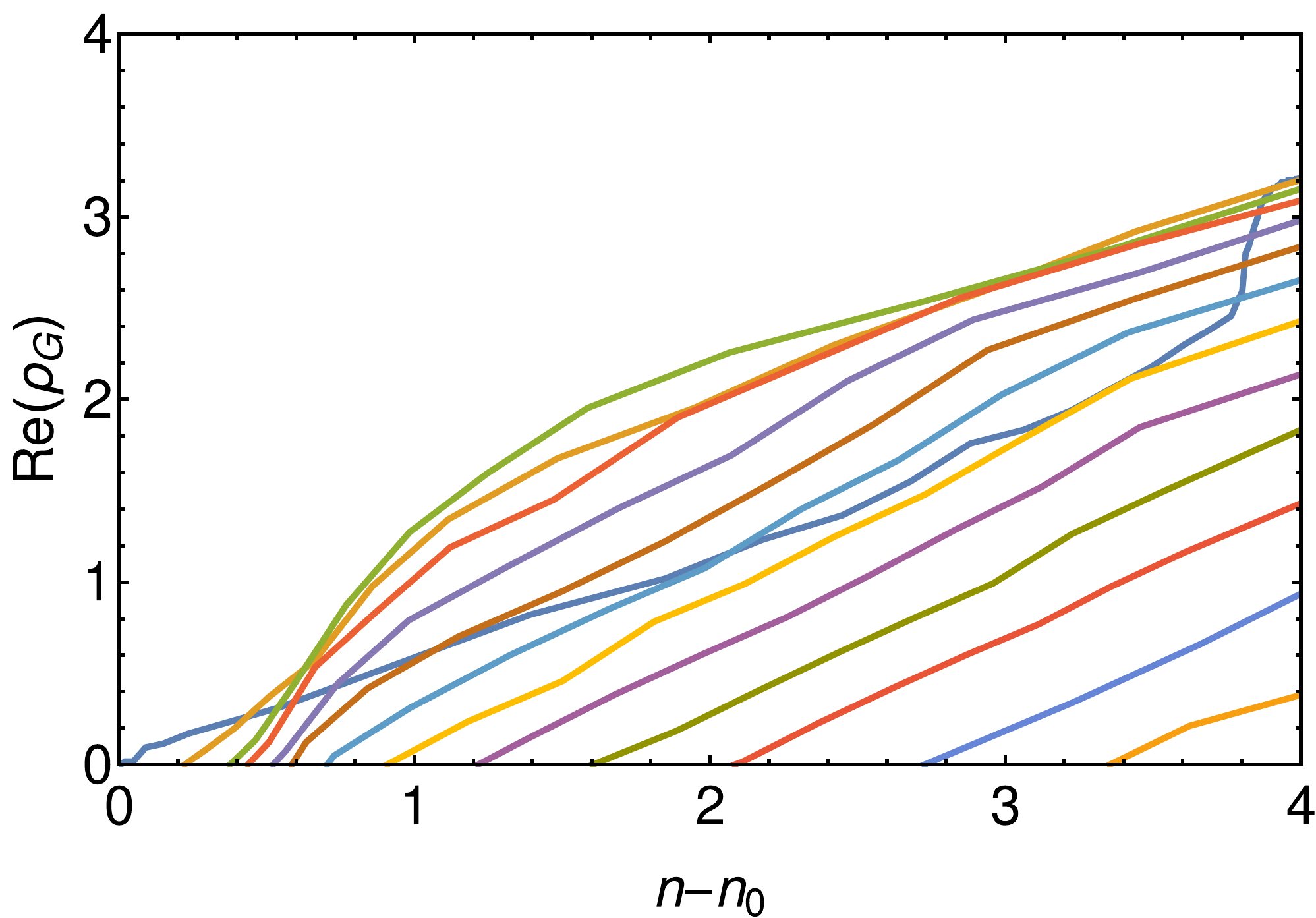} 
\end{center}
\caption{Dependence of the real part of $\rho_G$ on band filling for a twisted bilayer graphene with Hartree potentials described by different values of the amplitude $V_H$. The band filling is normalized such that $n_0$ corresponds to the half filled case, and $n-n_0=4$ to the case where the four upper bands are filled. We use values of $V_H$ from $0$ to $24$ meV.}
\label{fig:comp}
\end{figure}

As we increase the Hartree potential, the energies at the $K, K'$ and $M$ points are raised in comparison to those at the $\Gamma$ point, and the upper band becomes flatter (as can be seen in Figs.~\ref{fig:bandPT}, this effect is largely described by first order perturbation theory). We show a selection of bands for a few choices of the Hartree potential in  Fig.~\ref{fig:comp_hartree}. The Hartree potential is determined by the dimensionless parameter $\rho_G$, which by definition vanishes at the neutrality point. Its value shows an approximately linear a dependence on the fractional band filling, $n$, $\rho_G \approx A ( V_H ) (n-n_0(V_H))$, as can be seen in Fig.~\ref{fig:comp}.

For $\epsilon = 4$ and $L_M \approx d / \theta \approx 13$ nm, we obtain $V_0 \approx 28$ meV. Using this value, the solution to  \eqref{vh_selfc} is shown in Fig.\ \ref{fig:compSC}. The value of $V_H ( n )$ grows approximately linearly with $n$.
 The value of $V_H ( n )$ changes from 0 at the neutrality point, to $V_H ( 4 ) \approx 22\,\text{meV}$ when all the low energy bands are occupied.

{\it Exchange potential.}
We now consider the Fock part of the Hartree-Fock approximation. Exchange coupling takes place between states in  the same valley with identical spin. The exchange potential plays an important role in monolayer graphene, as it is the origin of the renormalization of the Fermi velocity \cite{GGV94,Eetal11}. Unlike the case of the Hartree term considered above, the effect of the exchange potential will lead to non-trivial changes in the bands at the neutrality point.  The small value of the Fermi velocity, $\tilde{v}_F$, in twisted bilayer graphene implies that the effective fine structure constant associated with the Dirac cones, $\alpha = e^2 / ( \epsilon \hbar \tilde{v}_F )$ is very large, $\alpha \sim 10^2$. The renormalization of $\tilde{v}_F$ has a small value for the high-energy cut-off, which is of the order of the bandwidth $W$. The renormalization of $\tilde{v}_F$ depends logarithmically on $W$ and linearly on $\alpha$, so that the corrections to $\tilde{v}_F$ will be important. The main effect will be a depletion of the density of states at the Dirac point.

Besides the renormalization of the Fermi velocity, the exchange potential leads to an overall widening of the bands. We can estimate this effect by considering the shift of the states at the band edges, located at the $\Gamma$ point. As the Coulomb potential is singular at zero momentum, the leading contribution to the shift of the states at the $\Gamma$ point comes from the interaction with occupied states with similar momenta. The occupied states at the bottom of the band have a similar internal structure to the neighboring ones, differing only in the crystal momentum. If we assume that this approximation is valid up to some cut-off in momentum space at a distance $\Lambda$ from the $\Gamma$ point, the energy shift has the value
\begin{align}
\delta \epsilon_\Gamma^{ex} &\approx - \frac{1}{4 \pi^2} \int^\Lambda 2 \pi \frac{e^2}{\epsilon k} k d k = - \frac{e^2 \Lambda}{2 \pi \epsilon}.
\end{align}
Assuming that $\Lambda \lesssim L_M^{-1}$, we find that the shift of the lower $\Gamma$ state, and the bandwidth, scales as $\delta \epsilon_\Gamma \approx e^2 / ( \epsilon L_M )$. We have analyzed numerically the decay of the overlap $S = | \langle \phi_{\Gamma} | \phi_{\vec{ k}} \rangle |^2$ as function of the distance between $\vec{ k}$ and $\Gamma$. Results are shown in Fig.~\ref{fig:overlap}. They show that $\Lambda \approx 0.1 \times ( 4 \pi ) / ( 3 L_M )$. We thus obtain
\begin{align}
\delta \epsilon_\Gamma^{ex} &\approx - 0.1 \,\frac{2 e^2}{3 \epsilon L_M}.
\end{align}
This energy is an approximation for  the increase in bandwidth. As the unoccupied state at the top of the conduction band has an internal structure which is orthogonal to the states at the lower $\Gamma$ point, it will not experience a significant exchange shift.

\begin{figure}
\begin{center}
\includegraphics[scale=0.50]{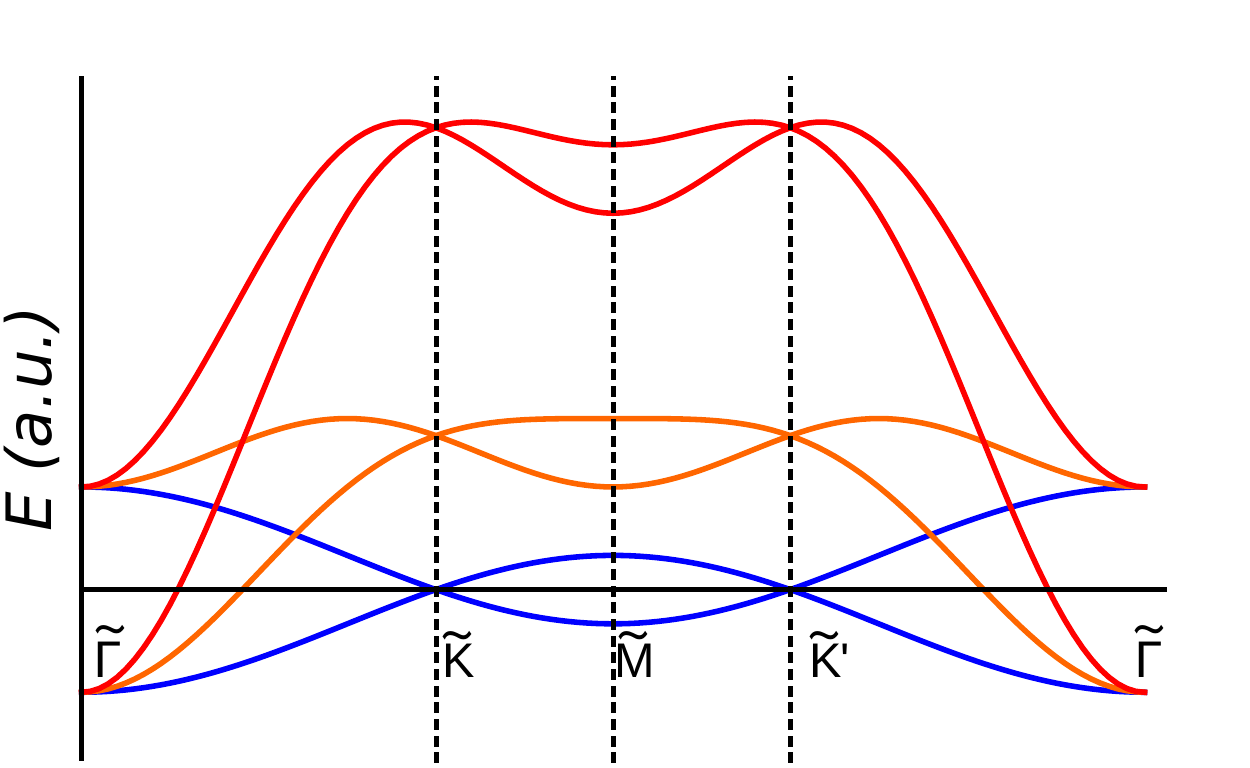} 
\end{center}
\caption{Band structure of a simple tight binding model based on the Wannier functions of twisted bilayer graphene, including the electrostatic potential. Black lines: bands at charge neutrality. Red and blue lines: bands as function of increasing Hartree potential.}
\label{fig:tight_binding}
\end{figure}

{\it Description of electrostatic effects in terms of local interactions.}
The Hartree potential significantly distorts the lowest band in a twisted bilayer graphene, but it leaves the bands which lie further away from the Dirac point mostly unchanged. Hence, it can be expected that the Hartree potential does not  significantly alter the Wannier wavefunctions for these bands. These Wannier functions have been extensively discussed, see, e.g., Refs.~\cite{PZVS18b,Ketal18b,KV18b}. For a given valley, the two Wannier wavefunctions can be approximated by functions centered in the $AB$ and $BA$ regions of the Moir\'e pattern. These regions form a honeycomb lattice. Each Wannier function has a three lobe structure with lobes peaked at the $AA$ regions. The Hartree potential can be projected onto these wavefunctions, and it can be written in terms of diagonal and off diagonal matrix elements. Each Wannier function is a four component spinor, with weight on both sublattices in each of the two layers, that is
\begin{align}
| m \rangle &= \sum_{i_S,i_V} \alpha_m^{i_S,i_V} | i_S , i_V \rangle,
\end{align}
where $\{ i_S,i_V \}$ label the sublattice and valley degrees of freedom. Each of the components has a different symmetry around the $AA$ lobes,  with angular momentum $\ell = -1 , 0 , 0 , 1$, see Ref.~\cite{Ketal18b}.  The Hartree potential is diagonal in sublattice, layer, and spin space. The matrix elements $\langle m | V_H | m \rangle$ where $m , n = 1 , \cdots , 6$ label the six sites around a given $AA$ region, can be written as
\begin{align}
\langle m | V_H | n \rangle &= \sum \left( \alpha_m^{i_S,i_V} \right)^* \alpha_n^{i_S,j_V} \langle i_S,i_V | \hat{V}_H | i_S,i_V \rangle.
\label{matrix}
\end{align}
The functions $m,n$  belong either to the $\bar{A}$ or the  $\bar{B}$ sublattice in the emergent honeycomb lattice with the Moir\'e pattern as a building block. We choose the  sites $m = 1 , 3 , 5$ to belong to the $\bar{A}$ sublattice, and the sites $m = 2 , 4 , 6$ belong to the $\bar{B}$ sublattice. The angular momentum of the different components of each Wannier function determines the phase of the amplitudes $\alpha_m^{i_S,i_V}$ in \eqref{matrix}. Matrix elements between functions which belong to different sublattices require the evaluation of integrals between states with angular momentum $\ell = \pm 1$ and states with $\ell = 0$. The phases shown in Ref.~\cite{Ketal18b} suggest that these elements vanish. The remaining matrix elements describe a second nearest neighbor hopping term in the Moir\'e lattice. Hence, we can add the Hartree potential to the simplified two-parameter tight binding model described in Ref.~\cite{Ketal18b} to obtain
\begin{align}
{\cal H}_{\text{local}} &= {\cal H}_0 + {\cal H}_H =  t_1 \sum_{\langle i , j \rangle} c^\dag_i c_j + i t_2 \sum_{\langle \langle i,j \rangle \rangle} c^\dag_i c_j + \nonumber \\ & \bar{V}_H \sum_{\langle \langle i,j \rangle \rangle , \{ i,j \} \in \{ \Bar{A} , \bar{B} \}} c^\dag_i c_j + \text{h.c.}\,,
\end{align}
where $t_1$ and $t_2$ are real-valued tight binding parameters, $\bar{V}_H$ parametrizes the Hartree potential,  $\langle i , j \rangle$ and $\langle \langle i,j \rangle \rangle$ are first and second nearest neighbors. This Hamiltonian ignores  a number of the hopping terms considered in Refs.~\cite{PZVS18b,Ketal18b,KV18b}, but the terms included are sufficient for a discussion focused on the role of the Hartree potential.
Results for different values of $\bar{V}_H$ are shown in Fig.~\ref{fig:tight_binding}. The bands are in reasonable agreement with the bands obtained using the continuum model, shown in Fig.~\ref{fig:comp_hartree}.

The description of the Hartree potential in terms of effective hopping parameters allows us to define an Wannier-based model for the long-range Coulomb interaction in graphene. The occupation of Wannier functions leads to charge accumulation at the $AA$ sites. The strength of the potential at a given site is proportional to the occupancy of the six Wannier functions centered at the neighboring $AB$ and $BA$ sites. The ensuing potential leads both to a change in the on-site energies, and to the creation of effective couplings between Wannier orbitals which are second nearest neighbors in the honeycomb lattice, but overlap in the same $AA$ region. We obtain the effective Hamiltonian
\begin{align}
{\cal H}_{\text{int}} &=  \sum_{m} \left[ V_{H 1} \left(  \sum_{i= 1 , \cdots , 6} c_{i,m}^\dag c_{i,m} \right)^2 + \right. \nonumber \\ & \quad\left.  V_{H 2} \left(  \sum_{i'= 1 , \cdots , 6} c_{i',m}^\dag c_{i',m} \right) \left(  \sum_{\langle \langle i , j  \rangle \rangle} c_{i,m}^\dag c_{j,m} \right) \right],
\label{hlocal}
\end{align}
where the sum over the $m$ index implies a sum over the centers of the hexagons in the honeycomb lattice, which define the $AA$ sites. The first term in ${\cal H}_{\text{int}}$ describes the local repulsion between charges placed at the $AA$ sites, and it has been discussed in Ref.~\cite{XLL18b}. The second term in ${\cal H}_{\text{int}}$ describes a hopping which depends on the charge state at the $AA$ regions. The description of the Hartree potential depends only on the Coulomb interaction, parametrized by $e^2/ ( \epsilon L_M )$. Hence, we expect that both terms in ${\cal H}_{\text{int}}$ scale as $V_{H1} , V_{H2} \propto e^2 / ( \epsilon L_M )$. The relation between $V_{H1}$ and $V_{H2}$ can be inferred from the band structures shown in Fig.~\ref{fig:comp_hartree}. The Hartree potential used to calculate the bands, $V_H$, gives the scale of the on-site term in ${\cal H}_{\text{int}}$ in \eqref{hlocal}, parametrized by $V_{H1}$, while the relative shift between the $\Gamma$ and $K$ points is due to the assisted hopping term, $V_{H2}$. The difference in the Hartree potential between the $AB$ and $AA$ regions in the Moir\'e cell is $9  V_H$, which gives an upper bound to the on site energy described by $V_{H1} , V_{H1} \lesssim ( 9  V_H ) / 3$ (note that the weight of a Wannier on a given $AA$ region is $1/3$). On the other hand, the shift of the $K$ point with respect to the $\Gamma$ point induced by the hopping term in \eqref{hlocal} is $\epsilon_K - \epsilon_\Gamma = 9\, V_{H2}$. From the results in Fig.~\ref{fig:comp_hartree}, we conclude that $V_{H2} \sim V_{H1}$.

{\it Electron assisted hopping and superconductivity.}
The existence of assisted hopping terms\footnote{After the submission of this manuscript Ref.~\cite{KV18c} was posted, also analyzing assisted hopping terms in graphene.} in the effective Hamiltonian of twisted bilayer graphene seems natural, as different Wannier functions overlap at regions where charging effects are maximal \cite{PZVS18b,Ketal18b,KV18b}. The effect of assisted hopping interactions in high-$T_c$ superconductivity has previously been studied in Refs.~\cite{HM90,MH90}. The most likely instability favored by such an interaction is superconductivity, since i) the effective interaction is attractive for some range of fillings, and ii) an assisted hopping term does not lead to many of the other typical broken-symmetry phases, such as a magnetic state, or a charge density wave.

\begin{figure}
\begin{center}
\begin{tabular}{ccc}
\includegraphics[scale=0.2]{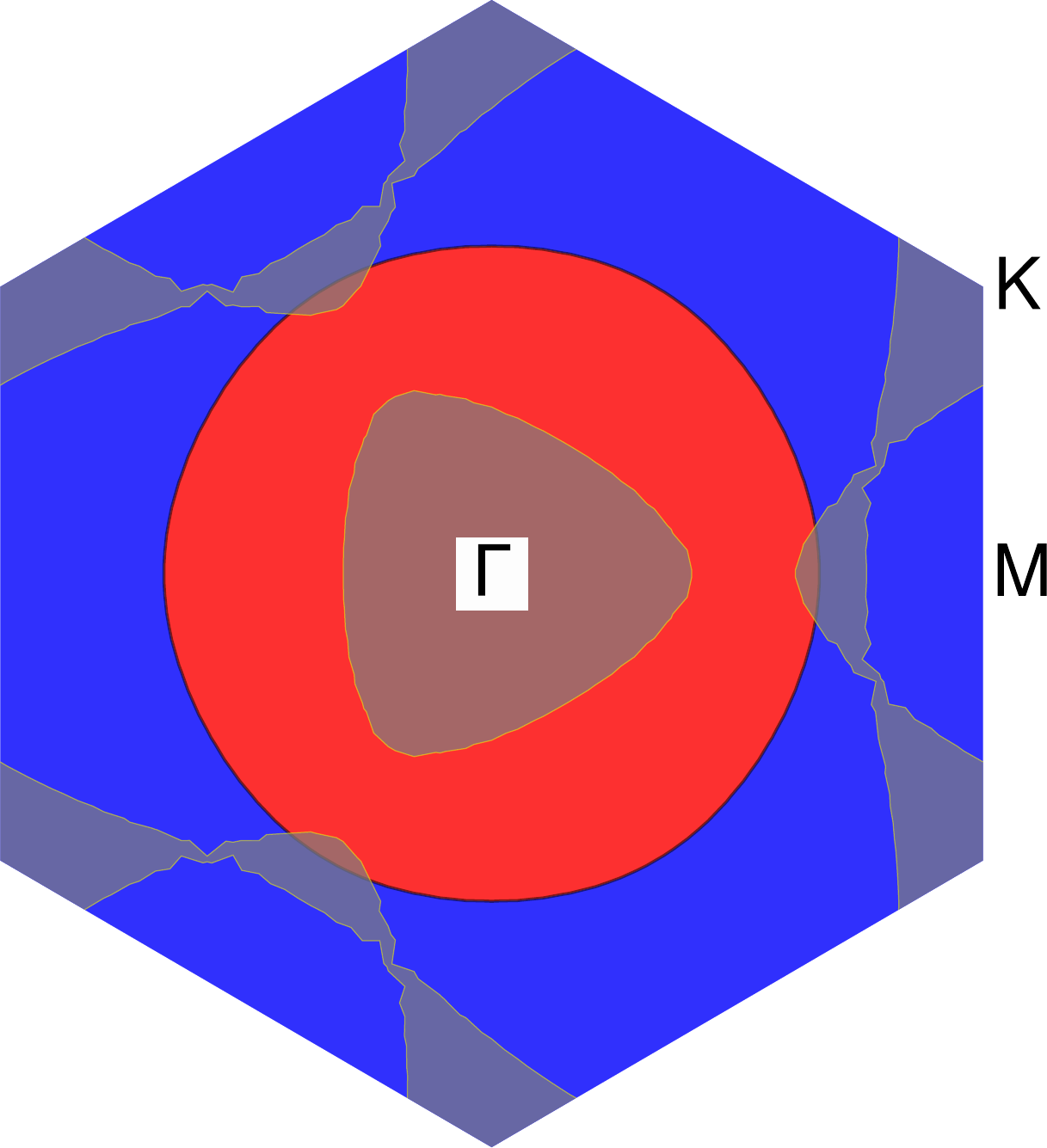} &
\includegraphics[scale=0.2]{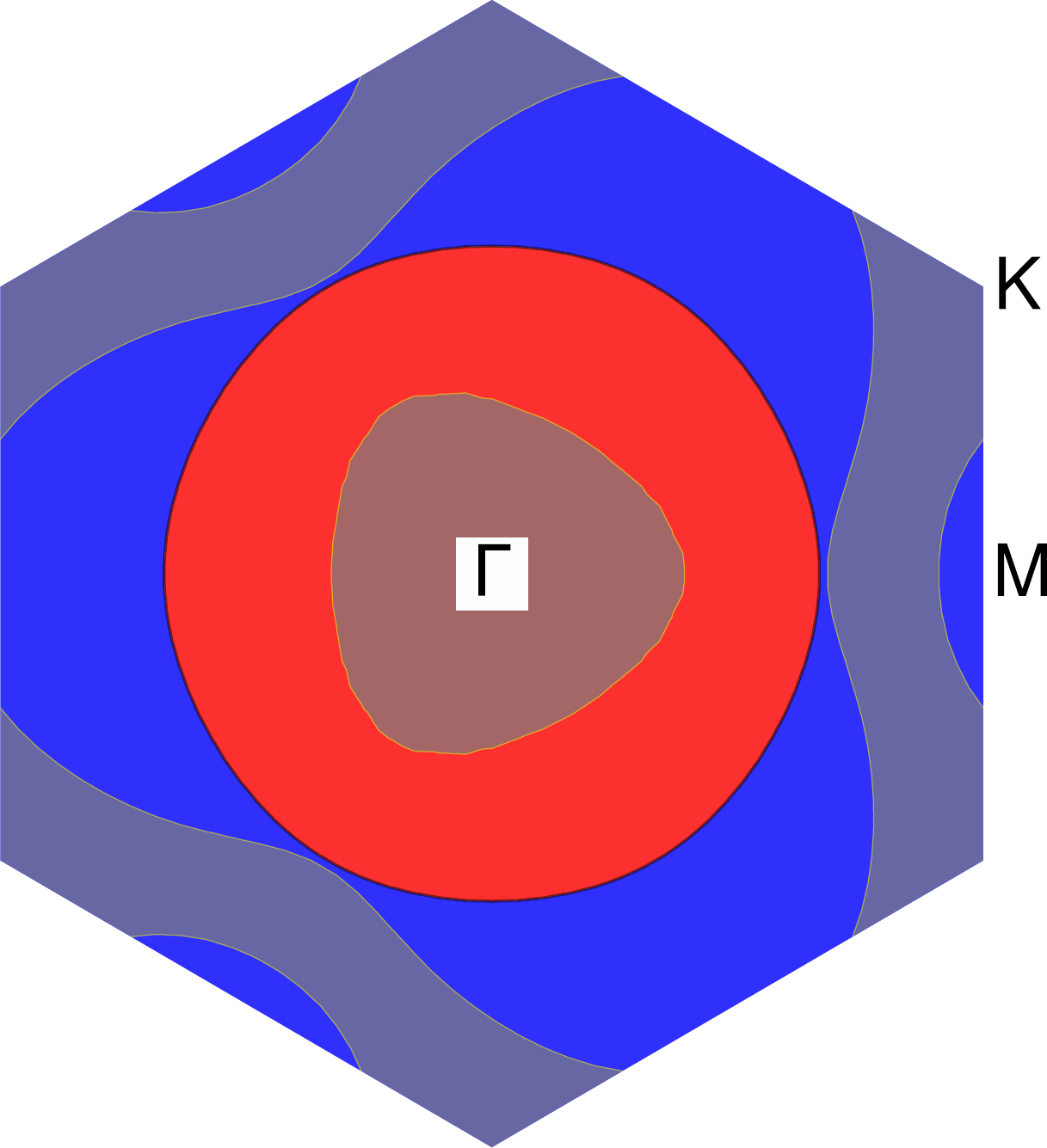} &
\includegraphics[scale=0.2]{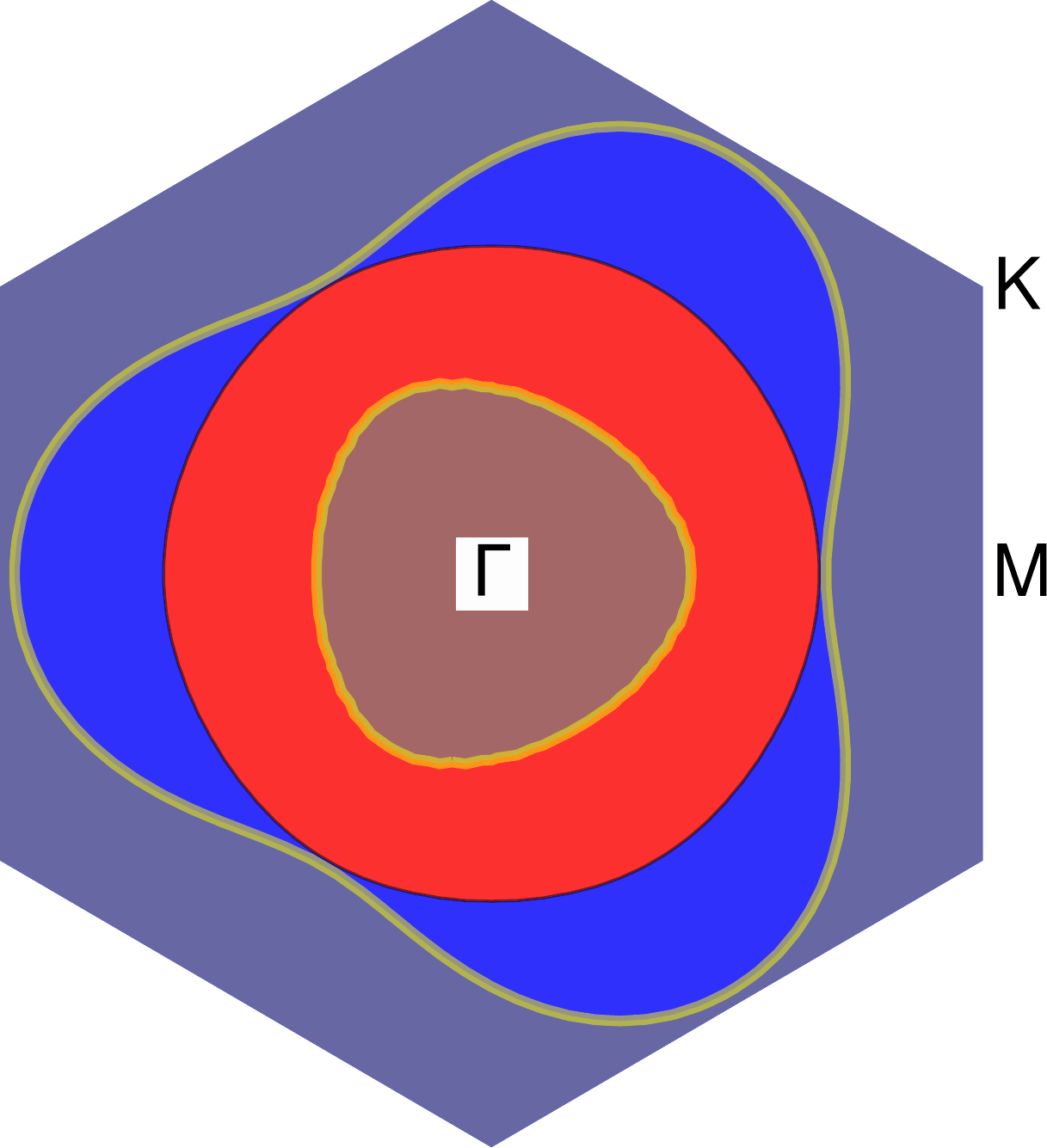} \\
\includegraphics[scale=0.2]{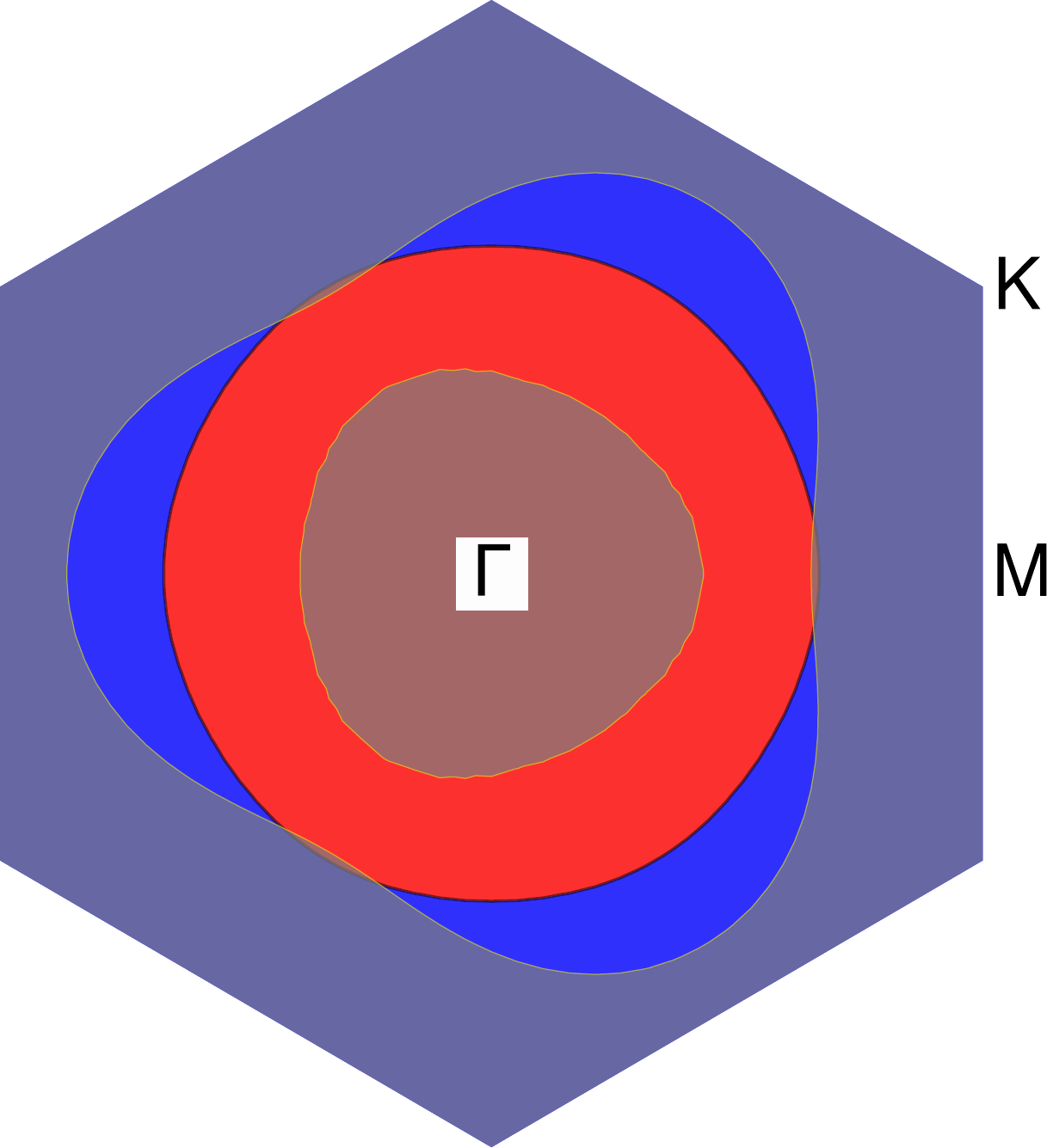}&
\includegraphics[scale=0.2]{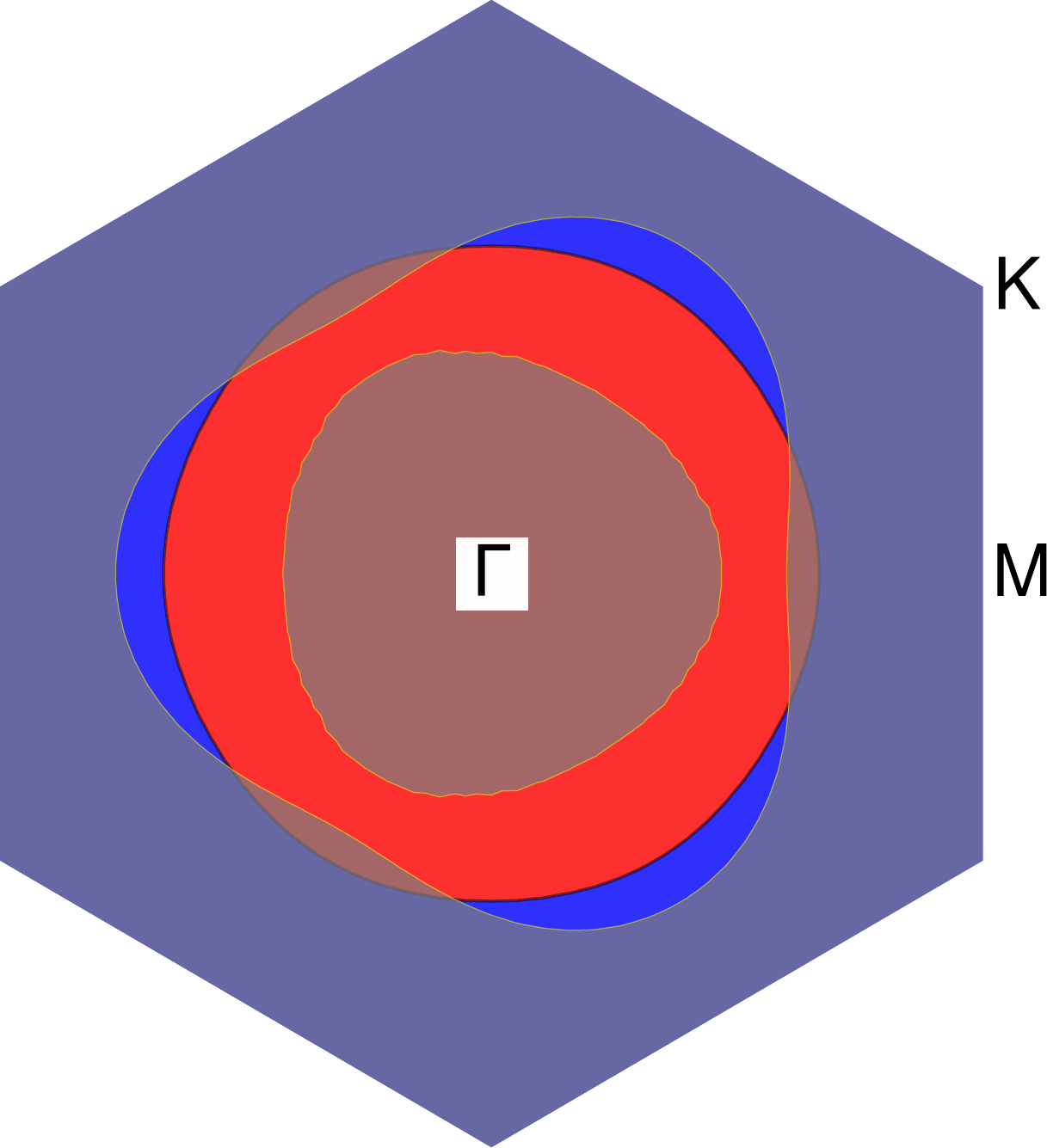} & 
\includegraphics[scale=0.2]{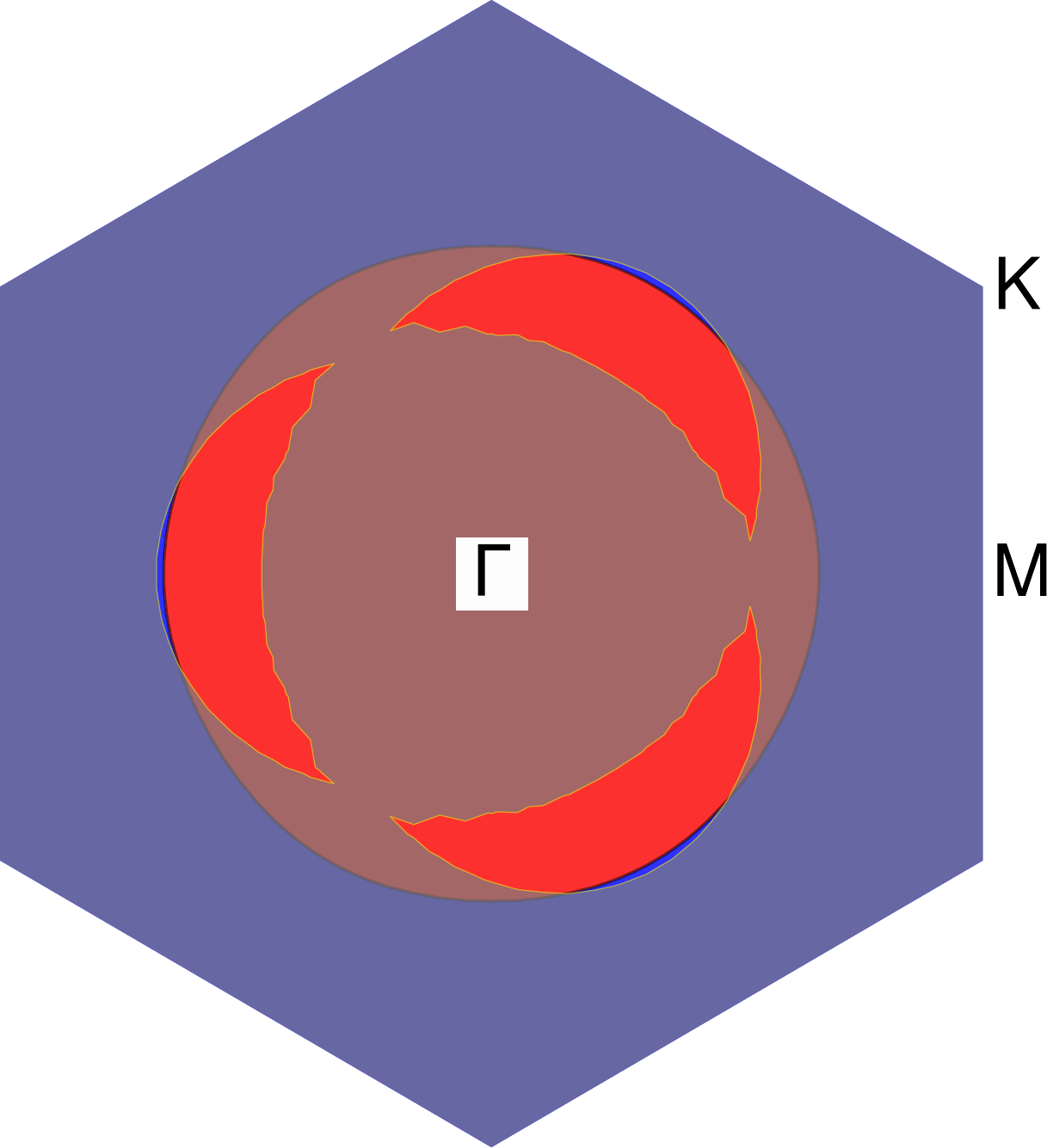}\\
\end{tabular}
\end{center}
\caption{Self consistent Fermi surface for different band fillings. The magnitude of $n$ gives the fraction of bands filled. The gray areas denote the filled states. The value $n = 0$ denotes the neutrality point, and $n = 4$ describes four filled bands. The function $g ( \vec{ k} )$ in \eqref{func_sc} is negative in the blue region, where superconductivity is favored. Top, from left to right: $ n = 1 , n = 1.5, n = 2$,  Bottom: $n=2.5, n = 3 , n = 3.5$. The Fermi surfaces are calculated using the full continuum Hamiltonian, including the Hartree term.}
\label{fig:superconductivity}
\end{figure}

We analyze the effect on superconductivity of the interaction term in \eqref{hlocal} using the Bogoliubov-de Gennes approximation. For simplicity, we consider here only superconducting states with spin zero and invariant under time reversal symmetry, although other phases are possible. The competition between the repulsion described by the term proportional to $V_{H1}$ and the assisted hopping term, $V_{H2}$ can lead to many superconducting phases.  The local description derived from \eqref{hlocal} is not strictly necessary for this step, see the Supplementary Information, Section 5. The electron and hole Hamiltonians are described in momentum space by two $2 \times 2$ matrices, one for electrons with a given spin and valley index, and another for holes with opposite spin and valley index. The mean field decoupling of the interaction term leads to
\begin{align}
{\cal H}_{\text{B-dG}} &= \left( {\cal H}_{mf} - \epsilon_F {\cal I}_\sigma \right) \tau_z + \Delta_1 ( \vec{ k} ) V_{H1}  {\cal I}_\sigma \tau_x + \nonumber \\ 
&\quad  \left[ \Delta_2 ( \vec{ k} ) \left( 1 + e^{i \vec{ k}\cdot \tilde{\vec{a}}_1} + e^{i \vec{ k} \cdot\tilde{\vec{a}}_2} \right) + \right. \nonumber \\
&\left. \quad\Delta_3 ( \vec{ k} ) \left( e^{i \vec{ k}\cdot ( - \tilde{\vec{a}}_1 - \tilde{\vec{a}}_2 )} + e^{i \vec{ k}\cdot ( - \tilde{\vec{a}}_1 + \tilde{\vec{a}}_2 )} + e^{i \vec{ k} \cdot( \tilde{\vec{a}}_1 - \tilde{\vec{a}}_2 )} \right) \right] \times
\nonumber\\ 
&\qquad V_{H1} \sigma_x \tau_x  +  \nonumber \\ & 
\Delta_4 ( \vec{ k} ) V_{H2} {\cal I}_\sigma \tau_x\,+ 
\Delta_5 ( \vec{ k} ) V_{H2} \sigma_x \tau_x\,.
\label{hamilbdg}
\end{align}
Here ${\cal I}_\sigma$ and $\sigma_x$ are the identity matrix and a Pauli matrix in sublattice space, $\tau_x$ and $\tau_z$ operate on the electron hole index, and $\epsilon_F$ is the Fermi energy.  The vectors $\tilde{\vec{a}}_i$ are the lattice vectors of the honeycomb lattice.  The values of $\Delta_1 ( \vec{\bf k} ) , \cdots , \Delta_5 ( \vec{\bf k} )$ have to be determined self consistently, see \eqref{delta} in the Supplementary Information, Section 6. 

The existence of superconductivity is determined by the functions  $f ( \vec{ k} )$ and in $g ( \vec{ k} )$ defined in \eqref{func_sc} in the Supplementary Information, Section 6.  The function $f ( \vec{ k} )$ is always positive. The terms in \eqref{hamilbdg} proportional to $V_{H1}$ describe a repulsive interaction which cannot lead to superconductivity in an isotropic s-wave like channel. This repulsive interaction is somewhat underestimated, as interactions over length scales longer than $L_M$ are ignored (see, however, the estimate of the screening length in the Supplementary Information, Section 5). On the other hand, the function $g ( \vec{ k} )$ is positive near the $\Gamma$ point, and negative in a large region near the edges of the Brillouin Zone. If $g ( \vec{\bf k} ) < 0$ superconductivity becomes possible. Fermi surfaces for different fillings are shown in Fig.~\ref{fig:superconductivity}, and compared with the region in the Brillouin Zone where $g ( \vec{ k} ) < 0$. 

The results in Fig.~\ref{fig:superconductivity} show that typically there are two pockets at the Fermi surface, and that, for a large range of fillings, superconductivity is possible in one pocket but not in the other. As a result, the superconducting state will show two different gaps, or a gap-less pocket coexisting with a gapped one. The order of magnitude of the higher superconducting gap will be $\Delta_{sc} \sim W e^{- ( W L_M  \epsilon ) / e^2}$, where $W$ is the bandwidth.  Repulsive interactions at the atomic scale not considered here, such as an on site Hubbard term, can suppress the superconducting phase at integer fillings, see Ref.~\cite{GLGS17}.

{\it Conclusions.} 
The occupation of the low energy bands in a twisted bilayer graphene leads to inhomogeneous electrostatic potentials of magnitude of order of $e^2 / ( \epsilon L_M )$, where $L_M$ is the Moir\'e unit length. This estimate is comparable, or larger, than the width of the band. 

Electrostatic effects, induced by charging the system away from the neutrality point, distort the bands significantly.  The  states at the edges of the Brillouin Zone, at $K$ and $M$,  are shifted with respect to the states near the center of the Brillouin Zone, the point  $\Gamma$. The exchange term, on the other hand, leads to an increase in the bandwidth, approximately a fraction of $e^2 / ( \epsilon L_M )$.

The band distortion induced by the electrostatic potential can be described in terms of induced assisted hopping couplings. These terms fit naturally with the complex overlapping Wannier functions which give a local description of twisted bilayer graphene. Assisted hopping interactions favor generally superconductivity, and we explicitly show that $s$-wave pairing is possible at certain fillings.

\begin{acknowledgments}We would like to thank E. Bascones, M. J. Calder\'on, V. T. Phong, B. Amorim,  R. Miranda, and A. K. Geim for useful conversations.
This work was supported by the European Commission under the Graphene Flagship, contract CNECTICT-604391.
\end{acknowledgments}

\bibliography{moire_v3}
\newpage
\section*{Supplementary material}
\renewcommand\thefigure{S\arabic{figure}}    
\renewcommand{\theHfigure}{S\arabic{figure}}
\renewcommand\theequation{S\arabic{equation}} 
\renewcommand{\theHequation}{S.\arabic{equation}}
\setcounter{figure}{0}    
\setcounter{equation}{0}    
{\em Exchange interactions near the $\Gamma$ point.}
\begin{figure}[H]
\begin{center}
\includegraphics[width=3in]{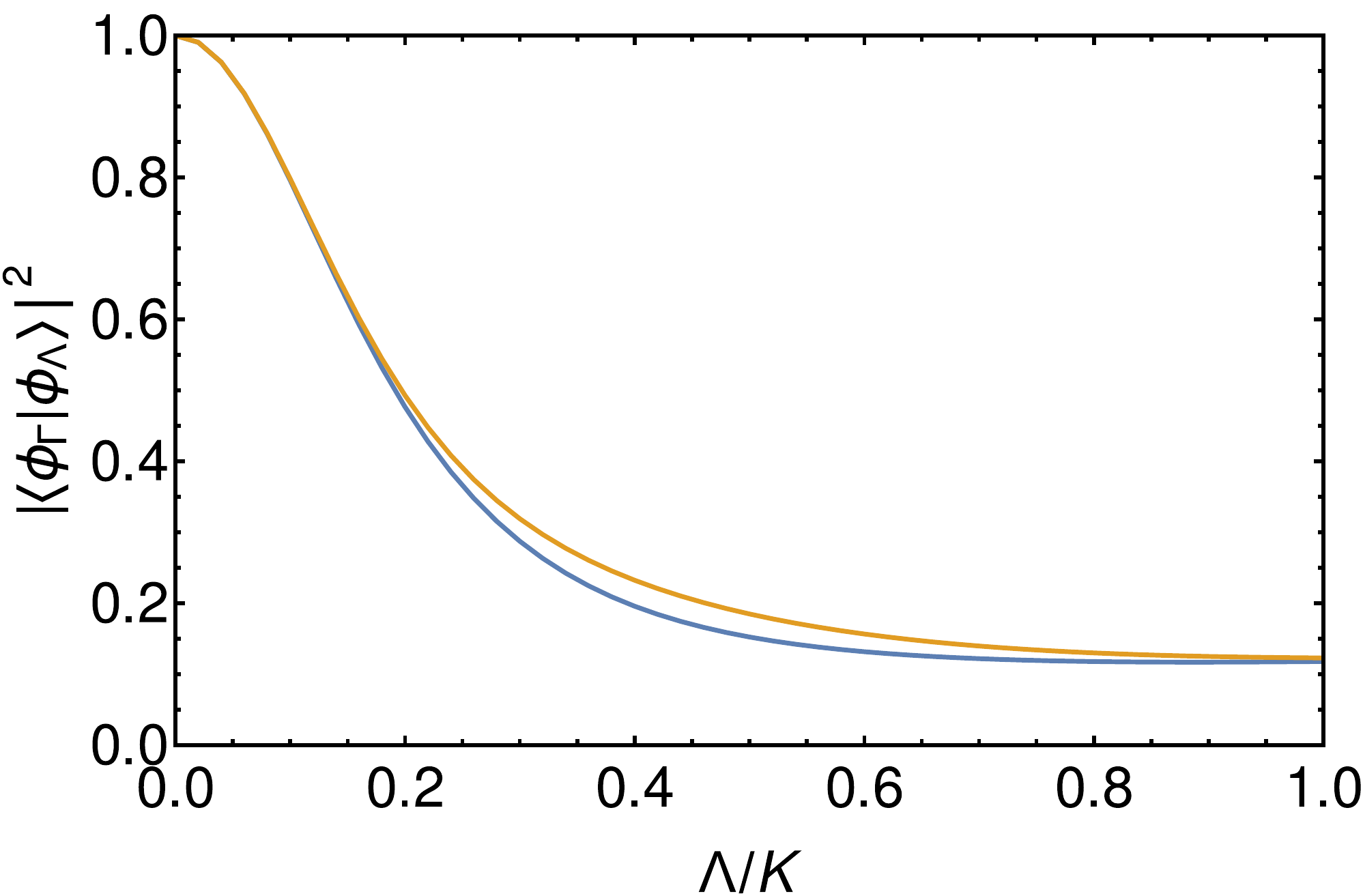} 
\end{center}
\caption{Dependence of the overlap between the wavefunctions of states at $\Gamma$ and at a momentum $\vec{\Lambda}$ along the $\Gamma - K$ direction. (Blue line for the top layer,  and yellow for the bottom layer.) The momentum at the $\Gamma$ point is zero, and $K$ is the magnitude of the momentum at the $K$ point.}
\label{fig:overlap}
\end{figure}

\begin{widetext}
{\em Deformation of the band structure induced by the Hartree potential.}

\begin{figure*}[h]
\begin{center}\includegraphics[width=5in]{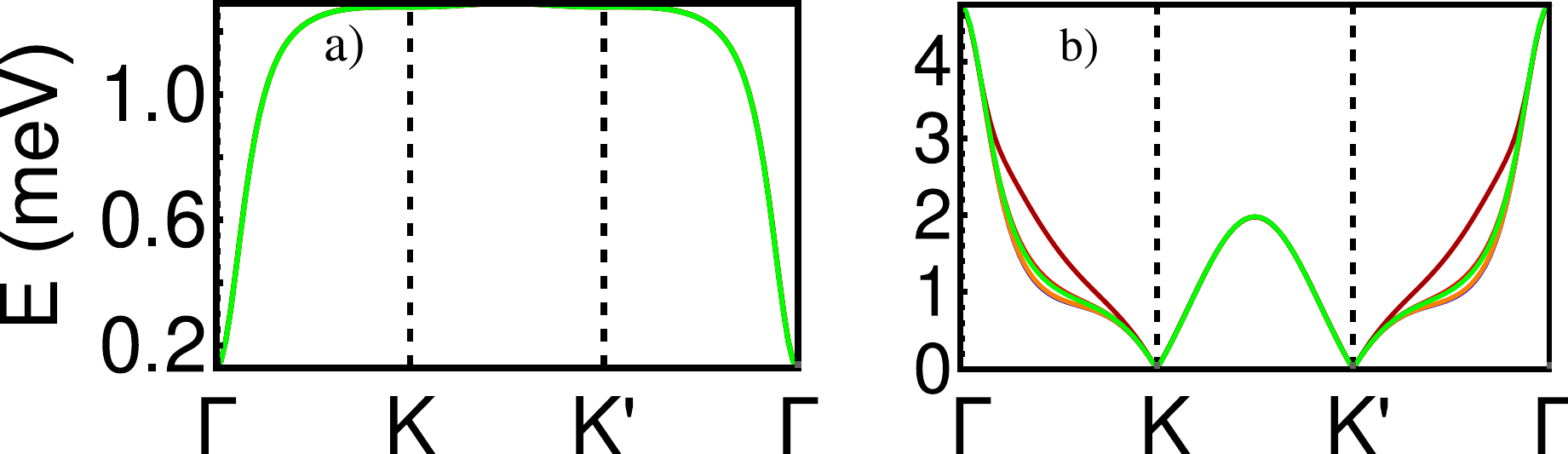}\end{center}
\caption{a) average energy shift of the bands in Fig.~\ref{fig:comp_hartree} divided by the value of $V_H$; b) band splitting  as  a function of $V_H$. Colors as in Fig.~\ref{fig:comp_hartree}.}\label{fig:bandPT}
\end{figure*}
\end{widetext}
The results in Fig.~\ref{fig:bandPT} show that the effect of the Hartree potential is a $\vec{\bf k}$ dependent shift of similar magnitude for the two bands, while the separation between them changes little. This result supports the description of the bands by an effective $2 \times 2$ Hamiltonian where the effect of the Hartree potential only enters as a $\vec{\bf k}$ dependent term in the diagonal elements. This term can be expressed as a sum of intra-sublattice hopping terms, see \eqref{hlocal}. 

{\em Fourier components of the charge density. Antisymmetric part.}

\begin{figure}[H]
\begin{center}
\includegraphics[width=2.5in]{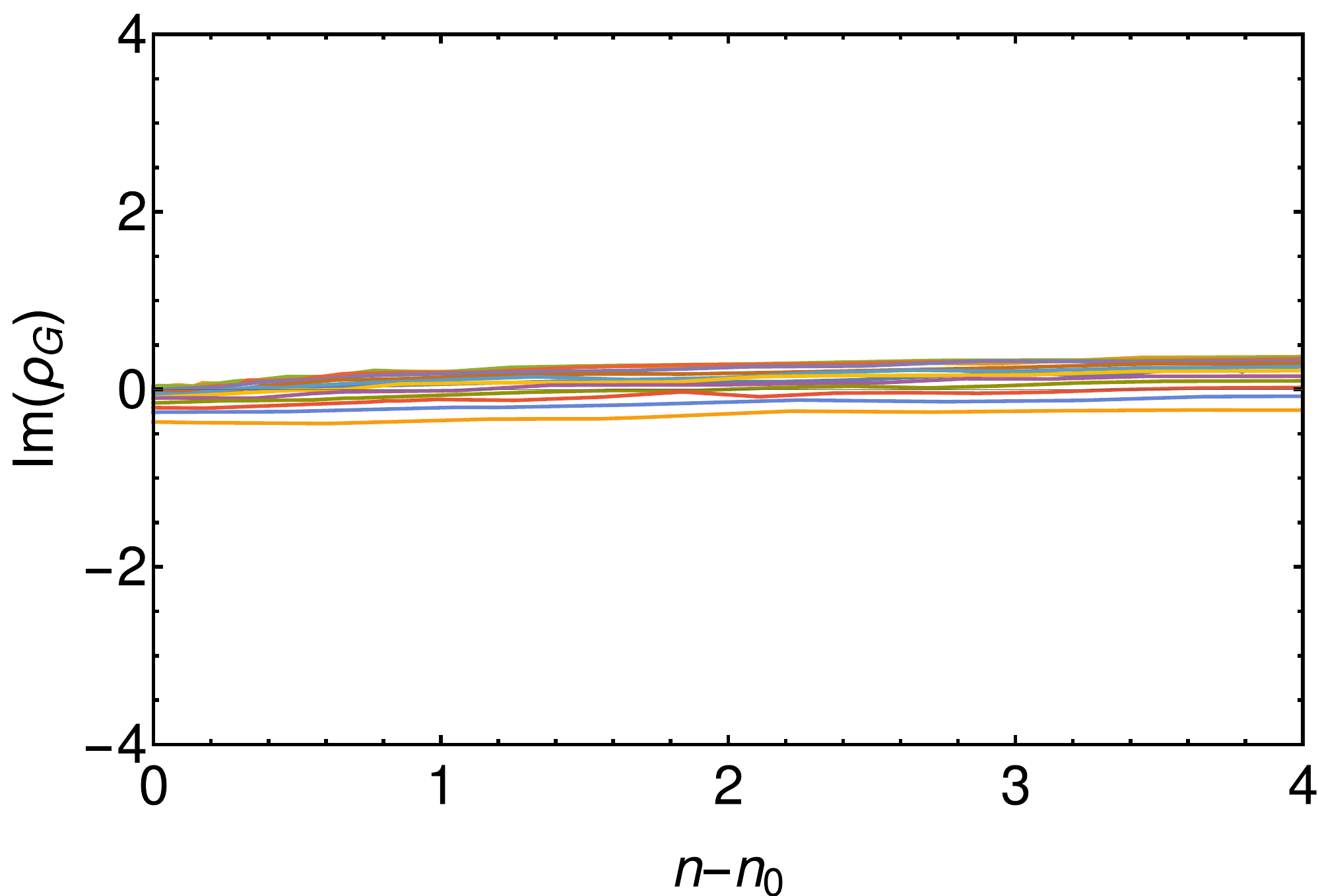} 
\end{center}
\caption{Dependence of the imaginary part of $\rho_G$ on band filling for a twisted bilayer graphene with Hartree potentials described by different values of the amplitude $V_H$. The band filling is normalized such that $n_0$ corresponds to the half filled case, and $n-n_0=4$ to the case when the four upper bands, described by their spin and valley index are filled. Colors and scale match those for the real parts in Fig.~\ref{fig:comp}.}
\label{fig:comp2}
\end{figure}

The imaginary part in $\rho_G$ reflects the difference in the effective interlayer hopping between the points on the same, e.g. $A\rightarrow A$, and different, e.g. $A\rightarrow B$,  sublattices. Following Ref.\ \cite{Ketal18b} we use $t_{AA} = t_{BB} = 79.7$ meV, and $t_{AB} = t_{BA} = 97.5$ meV. The Fermi velocity of graphene is $\hbar v_F = 5.5$ eV\AA.

{\em Details of the calculation of the Hartree potential.}

\begin{figure}[H]
\begin{center}
\includegraphics[width=2.5in]{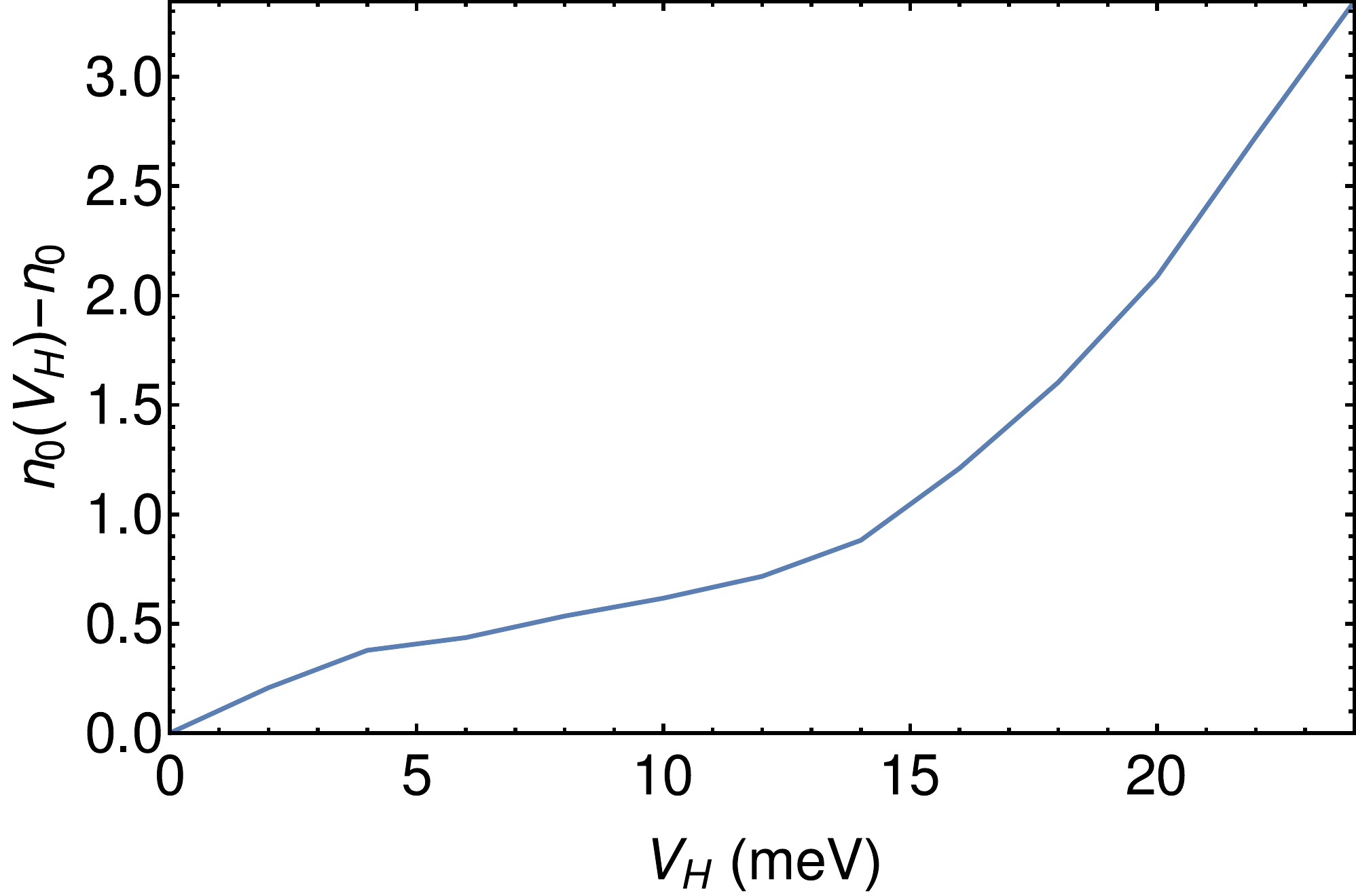} 
\end{center}
\caption{Zero crossings in Fig.~\ref{fig:comp} as a function of $V_H$; these we approximate as being linear, see the main tex.}
\label{fig:compzero}
\end{figure}

\begin{figure}[H]
\begin{center}
\includegraphics[width=2.5in]{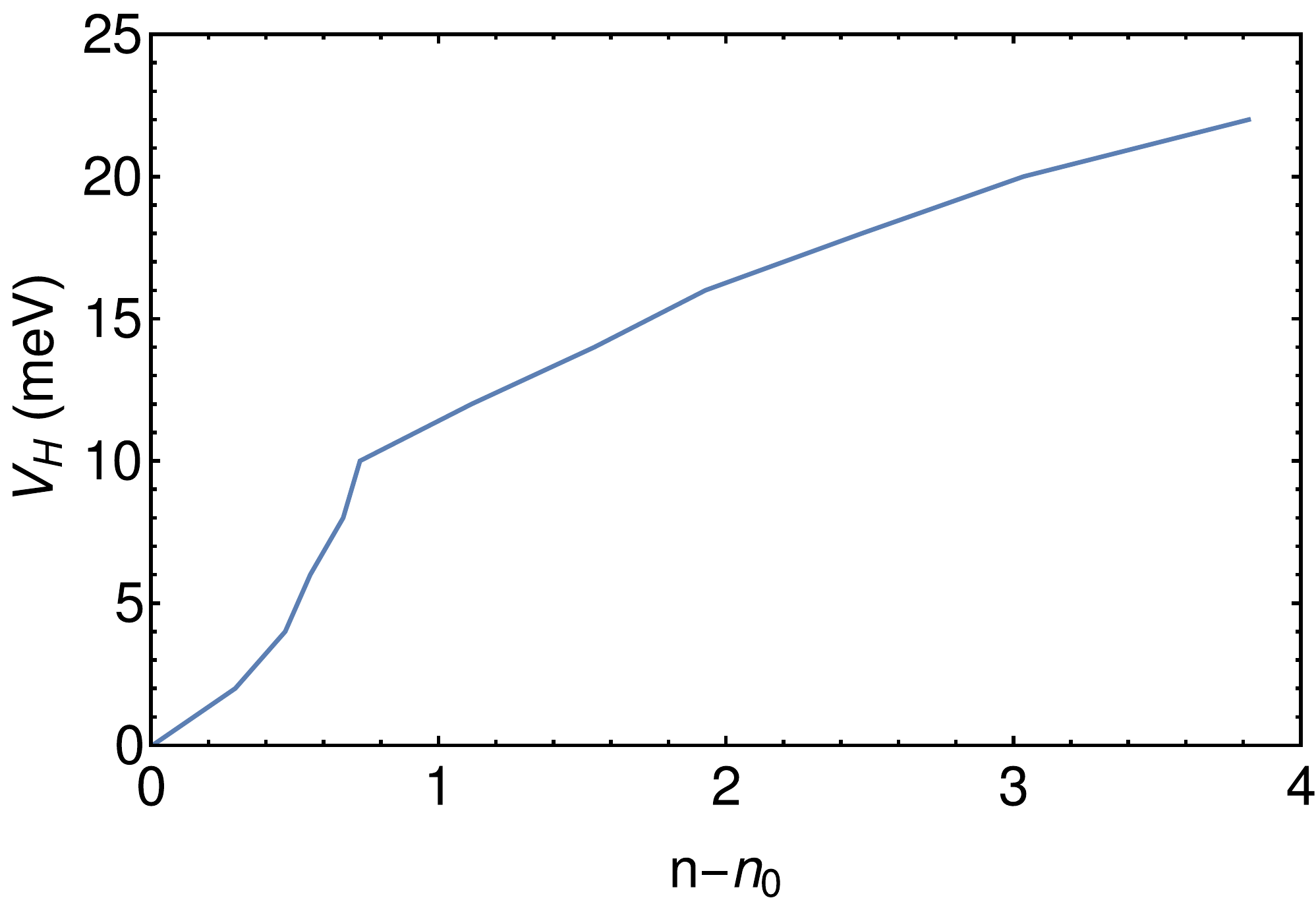} 
\end{center}
\caption{The self-consistent value of $V_H$  as a function of filling, using the value $V_0=28\,\text{meV}$. Cf.~Fig.~\ref{fig:comp}.}
\label{fig:compSC}
\end{figure}

{\em Definition of Wannier functions.}
The analysis of the effective interactions induced by charging effects, carried out in  of the main text, relies on the definition of localized Wannier functions. We use the description given  in Ref.~\cite{Ketal18b} of the Wannier functions for the continuum model of the twisted graphene layer. This description is consistent with similar work reported in\cite{PZVS18b,KV18}. 
The definition of a localized set of Wannier functions with all the symmetries of the underlying honeycomb lattice encounters difficulties similar to those found when studying surface states in topological insulators\cite{PZVS18b,ZPVS18,PWV17}. The use of the Wannier functions described in\cite{Ketal18b,KV18b} requires a fine tuning of the hopping parameters, but it also provides an intuitive framework where the new interactions can be identified. 

{\em Interactions in the weak coupling limit.}
The local description of the interactions induced by the Coulomb potential given in Eq.~(8) of the main text gives an intuitive picture of the physics. The analysis of superconducting phases which flows can be carried out independently of this local analysis. The interaction described in that equation can also be represented in Fourier space as
\begin{align}
\tilde{\cal H}_{int} &= \sum_{\vec{\bf k}} \hat{\rho}_{\vec{\bf k}} \frac{ ( 2 \pi e^2}{\epsilon_{FT} ( \vec{\bf k} )} \hat{\rho}_{- \vec{\bf k}}\,,
\label{hamilfourier}
\end{align}
where
\begin{align}
\epsilon_{FT} ( \vec{\bf k} ) &= \epsilon + \frac{2 \pi e^2}{\epsilon | \vec{\bf k} |} D ( \epsilon_F )\,.
\label{epsilon_ft}
\end{align}
Here $\epsilon_{FT} ( \vec{\bf k} )$ is the Fermi-Thomas screening function and $D ( \epsilon_F )$ is the density of states at the Fermi energy,  calculated using the {\it self consistent} bands. The weak coupling analysis can be done by projecting the Hamiltonian in \eqref{hamilfourier} onto the states at the Fermi surface.

A Fermi-Thomas screening length, $\lambda_{FT}$, can be obtained from \eqref{epsilon_ft}
\begin{align}
\lambda_{FT} &= \frac{\epsilon^2}{2 \pi e^2 D ( \epsilon_F )}
\end{align}
If we assume that
\begin{align}
D ( \epsilon_F ) \approx \frac{1}{W L_M^2}
\end{align}
where $W \sim 1 - 5$ meV is the bandwidth, we obtain $\lambda_{FT} \lesssim L_M$, which justifies a posteriori the use of the local approximation implicit in Eq.~(8) of the main text.

{\em Self consistency of the Bogoliubov-de Gennes equations.}

The quantities $\Delta_1 ( \vec{ k} ) \cdots\Delta_5 ( \vec{ k} )$ in \eqref{hamilbdg} have to be determined self consistently:
\begin{align}
\Delta_1 ( \vec{ k} ) &= V_{H1} \sum_{\vec{ k}'} f ( \vec{ k} - \vec{ k}' ) \langle  {\cal I}_\sigma  \tau_x \rangle_{\vec{ k}'} ,\nonumber \\
\Delta_2 ( \vec{ k} ) &= \frac{V_{H1}}{3}   \sum_{\vec{ k}'}  f ( \vec{ k} - \vec{ k}' ) \left( 1 + e^{- i  \vec{ k}'\cdot \tilde{\vec{a}}_1} + e^{- i \vec{ k}'\cdot \tilde{\vec{a}}_2} \right) \langle \sigma_- \tau_x \rangle_{\vec{ k}'}, \nonumber \\ 
\Delta_3 ( \vec{ k} ) &= \frac{V_{H1}}{3}   \sum_{\vec{ k}'}  f ( \vec{ k} - \vec{ k}' ) \left(  e^{i \vec{ k}'\cdot (  \tilde{\vec{a}}_1 + \tilde{\vec{a}}_2 )} + \right. \nonumber
\\&\qquad\qquad\qquad
 \left. e^{i \vec{ k}'\cdot (  \tilde{\vec{a}}_1 - \tilde{\vec{a}}_2 )} + e^{i \vec{ k}' \cdot( - \tilde{\vec{a}}_1 + \tilde{\vec{a}}_2 )} \right) \langle \sigma_- \tau_x \rangle_{\vec{ k}'}, \nonumber \\ 
\Delta_4 ( \vec{ k} ) &= V_{H2} \sum_{\vec{ k}'} f ( \vec{ k} - \vec{ k}' ) \left[ g ( \vec{ k} ) + g ( \vec{ k}' ) \right] \langle {\cal I}_\sigma  \tau_x \rangle_{\vec{ k}'}, \nonumber \\ 
\Delta_5 ( \vec{ k} ) &= V_{H2} \sum_{\vec{ k}'} f ( \vec{ k} - \vec{ k}' ) \left[ g ( \vec{ k} ) + g ( \vec{ k}' ) \right] \langle \sigma_-  \tau_x \rangle_{\vec{ k}'}\,.
\label{delta}
\end{align}
The first three terms correspond to intra-sublattice nearest-neighbor, inter-sublattice nearest-neighbors and inter-sublattice next-nearest-neighbor density-density couplings, and the last two terms correspond to intra and inter sublattice terms associated to assisted hopping. The functions $f ( \vec{ k} )$ and $g ( \vec{ k} )$ in \eqref{delta} are
\begin{align}
f ( \vec{ k} ) &= 3 + 2 \cos \left( k_x \right) + 4 \cos \left( \frac{k_x}{2} \right) \cos \left( \frac{\sqrt{3} k_y}{2} \right), \nonumber \\
g ( \vec{ k} ) &= 2 \cos \left( k_x \right) + 4 \cos \left( \frac{k_x}{2} \right) \cos \left( \frac{\sqrt{3} k_y}{2} \right).
\label{func_sc}
\end{align}

\end{document}